\documentclass[journal]{vgtc}                

\ifpdf
  \pdfoutput=1\relax                   
  \usepackage{graphicx}                
  \DeclareGraphicsExtensions{.pdf,.png,.jpg,.jpeg} 
\else
  \usepackage{graphicx}                
  \DeclareGraphicsExtensions{.eps}     
\fi%

\graphicspath{{figs/}{pictures/}{images/}{./}} 

\def\BibTeX{{\rm B\kern-.05em{\sc i\kern-.025em b}\kern-.08emT\kern-.1667em\lower.7ex\hbox{E}\kern-.125emX}}

\newcommand{\eat}[1]{}

\usepackage{mathptmx}                  
\usepackage{xcolor}
\usepackage[most]{tcolorbox}
\usepackage{latexsym}
\usepackage{amsfonts}
\usepackage{amsmath}
\usepackage{amssymb}
\usepackage{colortbl}
\usepackage{epsfig}
\usepackage{xspace}
\usepackage{graphicx}
\usepackage{subfigure}
\usepackage{paralist}
\usepackage{enumerate}
\usepackage{hyperref}
\usepackage{cleveref}
\usepackage[color,matrix,arrow,all]{xy}
\usepackage{comment}
\usepackage{booktabs}
\usepackage{balance}
\usepackage{stmaryrd}
\usepackage{pifont}
\usepackage{hhline}
\usepackage{listings}
\usepackage{array}
\usepackage{float}
\usepackage[flushleft]{threeparttable}
\usepackage{enumitem}
\usepackage{wrapfig}

\newcolumntype{L}[1]{>{\raggedright\let\newline\\\arraybackslash\hspace{0pt}}m{#1}}
\newcolumntype{C}[1]{>{\centering\let\newline\\\arraybackslash\hspace{0pt}}m{#1}}
\newcolumntype{R}[1]{>{\raggedleft\let\newline\\\arraybackslash\hspace{0pt}}m{#1}}

\usepackage{tikz}

\usepackage{epsfig}
\usepackage{multirow}
\usepackage{url}

\usepackage[export]{adjustbox}
\usepackage[noend]{algorithmic}
\usepackage{algorithm}

\newcommand*\circled[1]{\tikz[baseline=(char.base)]{
\node[shape=circle,fill=red,inner sep=1pt] (char) {\textcolor{white}{\small #1}};}}
\newcommand*\bluecircled[1]{\tikz[baseline=(char.base)]{
\node[shape=circle,fill=blue,inner sep=1pt] (char) {\textcolor{white}{\small #1}};}}

\makeatletter
    \newcommand\figcaption{\def\@captype{figure}\caption}
    \newcommand\tabcaption{\def\@captype{table}\caption}
\makeatother

\definecolor{yellow}{HTML}{FFFF00}
\definecolor{darkgreen}{rgb}{0.15,0.55,0.15}
\definecolor{darkblue}{rgb}{0.1,0.1,0.5}
\definecolor{blue}{rgb}{0.19,0.48,0.9}
\definecolor{darkgreen}{rgb}{0.15,0.55,0.15}
\definecolor{mred}{rgb}{.80,.12,.30}
\definecolor{grey}{rgb}{0.5,0.5,0.5}
\definecolor{Purple}{rgb}{.75,0,.85}
\definecolor{light-gray}{gray}{0.95}
\definecolor{mid-gray}{gray}{0.85}
\definecolor{darkred}{rgb}{0.7,0.25,0.25}

\newcommand{\red}[1]{\textcolor{red}{#1}}

\newtheorem{myExample}{Example}

\newtheorem{example}[myExample]{Example}

\newcommand{\stitle}[1]{\vspace{1ex}\noindent{\bf #1}}

\newtcbox{\mywboxtext}{on line,
colback=yellow,frame hidden,colframe=white,size=fbox,boxrule=0pt,fontupper=\color{red}}
\newcommand{\hl}[1]{\textrm{#1}}

\def\ojoin{\setbox0=\hbox{$\Join$}%
\rule[-0.05ex]{.27em}{.4pt}\llap{\rule[1.3ex]{.27em}{.4pt}}}
\def\leftouterjoin{\mathbin{\ojoin\mkern-5.8mu\Join}}

\def\fullouterjoin{\mathbin{\ojoin\mkern-5.8mu\Join\mkern-5.8mu\ojoin}}

\onlineid{0}
\vgtccategory{Research}
\vgtcpapertype{Representations \& Interaction}
\title{Extending the View Composition Algebra to Hierarchical Data}

\newcommand{\sys}[0]{\texttt{VCA}\xspace}
\newcommand{\sysh}[0]{\texttt{VCA}$^\texttt{H}$\xspace}

\author{Eugene Wu}
\authorfooter{
\item Eugene Wu is with Columbia University.  Email: ewu@cs.columbia.edu.
}

\shortauthortitle{Wu, E: Extending VCA to Hierarchical Data}


\vgtcinsertpkg

\abstract{Comparison is a core task in visual analysis.
Although there are numerous guidelines to help users design effective visualizations to aid known comparison tasks, there are few formalisms that define the semantics of comparison operations in a way that can serve as the basis for a grammar of comparison interactions.  
Recent work proposed a formalism called View Composition Algebra (VCA) that enables ad hoc comparisons between any combination of marks, trends, or charts in a visualization interface.

However, \sys limits comparisons to visual representations of data that have an identical schema, or where the schemas form a strict subset relationship (e.g., comparing {\it price per state} with {\it price}, but not with {\it price per county}).
In contrast, the majority of real-world data---temporal, geographical, organizational---are hierarchical.
To bridge this gap, this paper presents an extension to VCA (called \sysh) that enables ad hoc comparisons between visualizations of hierarchical data.
\sysh leverages known hierarchical relationships to enable ad hoc comparison of data at different hierarchical granularities.    We illustrate applications to spatial visualizations and Tableau visualizations.

}

\begin{document}
\maketitle

\section{Introduction}\label{s:intro}

Comparison is a core task in visual analysis.
Although there are numerous guidelines to help users design effective visualizations to aid known comparison tasks, there are few formalisms that define the semantics of comparison operations in a way that can serve as the basis for a grammar of comparison interactions.  

Recent work proposed a formalism called View Composition Algebra~\cite{vca} (VCA) that enables ad hoc comparisons between any combination of marks, trends, or charts in a visualization interface.
The key idea is to model a view $V = R(Q(D))$ as a thin visual mapping function over the result of a query $Q$.  $Q$ is responsible for all data processing, aggregation, and transformations.  In this way, composition operations over multiple views such as differencing and union can be defined as expressions over the underlying queries of the views.  

\begin{figure}[htpb]
  \centering
  \includegraphics[width=\columnwidth]{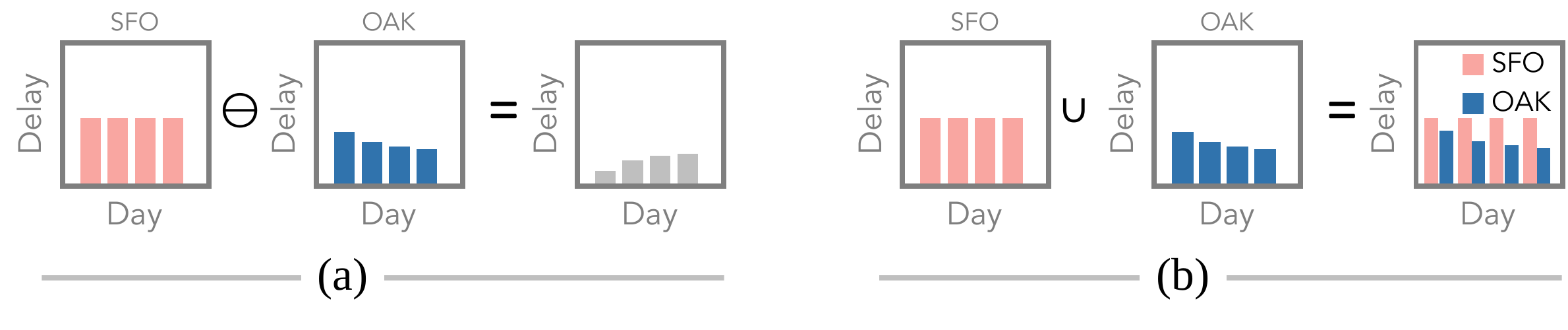}
  \caption{Example \sys operations.  (a) statistical composition computes the difference between two views, and (b) union composition juxtapose or superimposes data from both views.}
  \label{f:vca}
\end{figure}

For instance, \Cref{f:vca} illustrates the two core operators.  (a) computes the difference between SFO and OAK matching bars with the same days and subtracting each day's Oakland delay with the SFO delay on that day. (b) illustrates the union operation, which juxtaposes the bars for easier visua comparison.

However, \sys limits comparisons to views that both render the same data attributes, or where their data attributes have a subset relationship (e.g., comparing {\it price per state} with {\it price}).  
In contrast, the majority of real-world data are hierarchical, and uses may wish to compare e.g., {\it price per state in 2000} with {\it price per county in 2001}.

To bridge this gap, we extend VCA to enables ad hoc comparisons between visualizations of hierarchical data.  We call this extension \sysh.
\sysh leverages known hierarchical relationships to enable ad hoc comparison of data at different hierarchical granularities, and we base the formalisms on functional dependencies, a core concept in the relation model~\cite{codd1983relational}.   We also illustrate applications to the hierarchical language HIVE~\cite{slingsby2009configuring} and Tableau visualizations.

\section{Related Work}

\subsection{Visualization Languages for Hierarchical Data}

Hierarchical structures are integral to visualization algebras such as VizQL~\cite{hanrahan2006vizql,stolte03thesis}.   VizQL defines algebraic operators over data attributes to compose small-multiple display, which are manifested as interactions to drag-and-drop attributes onto x- and y-axis ``shelves''.    
The \texttt{cross} ($Qtr\times State$) operator creates one small multiple for each combination of quarter, state values.  
However, $\times$ is not appropriate when composing attributes within a hierarchy, because not all combinations of the attribute domains will exist in the dataset.  For instance, $Qtr\times Month$ will return $(Q1,Dec)$, which will never contain data.  
The hierarchy-aware \texttt{nest} ($Qtr/Month$) operator obeys the containment relationship, and ill only return pairs  where the month is in the appropriate quarter.  

Slingsby et al.~\cite{slingsby2009configuring} define a grammar for specify spatial layouts of hierarchical data.  Much like how graphical grammars map data attributes (e.g., day, cost) to a mark's visual attributes (e.g., x- and y-positions), HIVE proposes a visual mapping from attributes in a data hierarchy to levels in a spatial hierarchy.  In addition to this mapping, users also specify the desired layout algorithm at each level of the spatial hierarchy.  
For instance, using a simplified notation, \texttt{state(profit$\to$size)/year(profit$\to$size)} creates a nested tree map; the top level uses a space filling layout sized by price, and each rectangle is further subdivided by year and sized by that year's profits (\Cref{f:hive}).     HIVE can express a rich set of spatial visualizations, including calendars, tree maps, and cartograms. 
Similar ideas were explored in product plots~\cite{wickham2011product}, and probabilistic graphical grammars~\cite{pu2020probabilistic}.

Although these languages enable users to {\it create} visualizations that are aware of hierarchies, they
do not support ad hoc {\it comparisons across} different hierarchical levels.



\subsection{Composition in Visualization Grammars}

Graphical grammars, such as ggplot2~\cite{wickham2016ggplot2}, Vega-lite~\cite{Satyanarayan2017VegaLiteAG}, and VizQL~\cite{Stolte2000PolarisAS} model visualizations as mappings from data attributes to the visual attributes of the rendered marks.  These grammars implicitly perform data transformations, such as grouping and aggregation, based on the data types.   
For instance, VizQL is a table algebra to compose facetted, multi-layer visualizations, and used to navigate multi-dimensional data cubes~\cite{Gray2004DataCA,Stolte2003MultiscaleVU}.
\sys is compatible and composable with these grammars---\sys operators can take their views as input, and emits views as output.  \sys defines unambiguous data transformations so that interaction designers can focus on design choices.  

\subsection{Design Strategies for Comparison}

Javed and Elmqvist~\cite{Javed2012ExploringTD} propose a design space for composite visualizations, describe four visual composition designs (juxtapose, superpose with and without shared axes, and nested views).  Gleicher et al.~\cite{gleichercompare} characterize comparison by the target elements being compared, and the actions to compare them.  They similarly 
 propose juxtaposition, superposition, and explicit encoding as design strategies.  
\sysh goes beyond design guidelines, and defines formal composition rules and notions of safety.

\section{Preliminaries}

How to model hierarchical data so that an algebra can symbolically manipulate it?
This section presents a model for hierarchical data based on functional dependencies, and also introduces \sys.


\subsection{Functional Dependencies and Hierarchical Data}

SQL and most query languages are grounded in the relational model~\cite{codd1983relational}.  The relational model defines a database as a set of relations.   Each relation (table) consists of a set of tuples (records) that all adhere to a schema---a list of attribute, type pairs.   
Following VizQL~\cite{Stolte2000PolarisAS} and other visualization formalisms, we assume that a table $T$'s schema $A_T$ consists of dimensions (used for filtering, grouping, and database joins) and measures (used to compute statistics); without loss of generality, we will assume that each table contains a single measure. 

\stitle{Functional Dependencies:}
The relational model defines a functional dependency constraint (FD) $X\to Y$ as a set of attributes $X$ that {\it functionally determines} the set of attributes $Y$.  In other words, if two records have the same $X$ values, they must also have the same $Y$ values.  
For instance, $zip\to state$ says that two records with the same zipcode must have the same state.  
Foriegn key references are also a special case of functional dependencies because both values are equal, thus the dependency is in both directions.

\stitle{Hierarchies are FDs:}
Attributes in real-world data typically form a hierarchical structure. 
Hierarchies are used in multi-dimensional databases to drill-down or roll-up, and in visualization systems to zoom in or zoom out.  Visualization formalisms like VizQL~\cite{hanrahan2006vizql,stolte03thesis} rely on such hierarchies to define operators like {\it nest}.  

Hierarchies are a special case of functional dependencies.
When we say that \texttt{district} is a child of \texttt{state} in a hierarchy, 
is expressed as the FD \texttt{district}$\to$\texttt{state}.

Thus, we define a hierarchy $H = \{F_1,\ldots,F_n\}$
as a set of FDs whose edges form a directed acyclic graph.
In contrast to a colloquial hierarchy, where edges point from coarser to finer granularities (e.g., state to county), the edges in $H$ are reversed, and point from finer to coarser granularities.   

\begin{figure}[h]
  \centering
  \includegraphics[width=0.85\columnwidth]{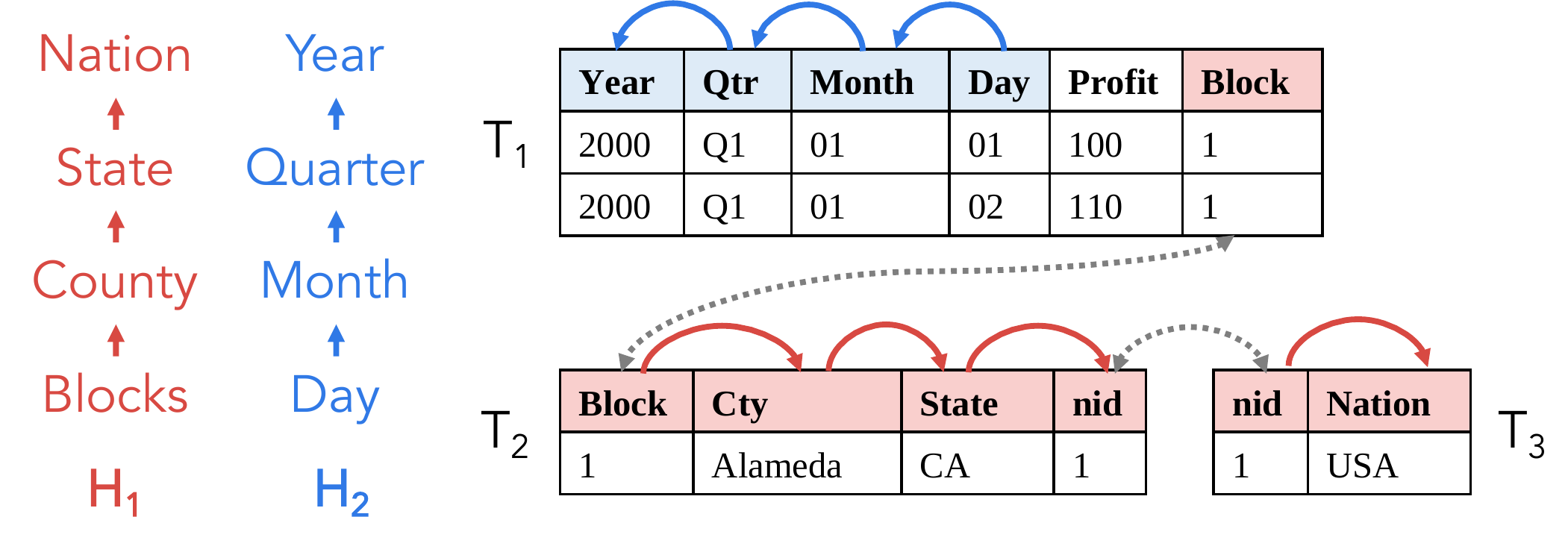}
  \caption{Tables with two example hierarchies; arrows are FDs.   }
  \label{f:hierarchy}
\end{figure}

For instance, \Cref{f:hierarchy} shows tables that contain hierarchical data.
$T_1$ stores each census block's daily \texttt{Profit}, and contains \texttt{Day}, \texttt{Month}, \texttt{Qtr}, and \texttt{Year} attributes.
$T_2$ stores the geographical data associated with each \texttt{block}, and can be joined with
$T_1$ on \texttt{block} (dotted line).
$T_3$ contains the \texttt{Nation} name, and joins with $T_2$ on \texttt{nid}.

%
%

\subsection{How VCA models Views}

\sys models a view as $V_i = R_i(T_i)$, where $T_i$ is a table (a raw table or a SQL query result) and $R_i$ is a visual encoding specification from a subset of $T$'s schema to visual attributes.  
In practice, $T_i = Q_i(D)$ is the output of a query $Q_i$ over an underlying database table $D$.  
Let $D(a_1,\ldots,a_n)$ contain $n$ attributes and $A_D = [a_1,\ldots,a_n, a_y]$ denote its schema, where $a_y$ is the measure, and the rest are dimensions.  Let $A_i$ be shorthand for $A_{R_i}$.

It is important that $Q$ encapsulates {\it all} computations and transformations needed to produce $T$, and that the renderer is only responsible for visual design---mapping data rows in $T$ to marks (or other objects) in the view and performing layout.  This is because the result of comparison operators depends on knowing {\it how} $T$ is computed.

\stitle{Queries:}
\sys focuses on {\it group-by aggregation queries}  
supported by most visual analysis systems~\cite{Stolte2000PolarisAS,dice2014,Gray2004DataCA,Chaudhuri1997AnOO,Pahins2017HashedcubesSL}.  These queries filter the input table using a predicate $p$, 
group records by a set of attributes $A_{gb}$, and compute an aggregated statistic $f_i(a_y)$:
  \begin{lstlisting}[
    caption={},
    captionpos=b,
    basicstyle=\small\ttfamily,
    frame=,
    mathescape=true,
    escapeinside={@}{!} ]
  $Q_i$ = SELECT @$A_{gb}$!, @$f(a_y)\to y$!  FROM D [WHERE @$p$!] GROUP BY @$A_{gb}$!
  \end{lstlisting}

To aid symbolic manipulation, \sys analyzes the equivalent relational algebra statements (\Cref{t:relalg} for full list):

  $$Q_i = \gamma_{A_{gb}, f(a_y)\to y}(\sigma_{p}(D))$$

\noindent $\sigma$ keeps rows that satisfy predicate $p$,
$\gamma$ groups rows by a subset of the dimensions $A_{gb}$ and computes  $f(a_y)$ for each group.

\stitle{Visual Mapping:}
Given the view's mark type $mark$, $R_i$ maps 
attributes in the query result to visual attributes  (e.g., x, y, color) valid for the mark type $K$. 
Let $A_{Q_i}$ be the attributes in $Q_i$, and $A_{mark}$ be the set of visual attributes for $mark$.  $R_i$ is defined as:
$$R_i = \{mark\to `mark'\} \cup \{ a_q\to a_v | a_q\in A_{Q_i} \land a_v\in A_{mark} \}$$
Each visual attribute can be referenced at most once, and not all query attributes need to be mapped.  

\begin{table}
    \centering \scriptsize
    \begin{tabular}{rl} 
      \textbf{Operator} & \textbf{Description} \\ 
       $\gamma_{A,f(a)}$  & Group by attributes $A$, and compute $f(a)$ for each group \\ 
       $\pi_{e_1\to a_1,\ldots}$  & Compute expressions $e_i$ and rename them as $a_i$.  \\
                        & $T.*$ copies all attributes from input table T. \\
       $\sigma_{p}$  & Filter records using boolean function $p(row)$ \\
       $S\bowtie_{A} T$  & Join $S$ and $T$ rows with the same attribute values in $A$  \\
    \end{tabular}
    \caption{Description of relational algebra operations}
    \label{t:relalg}
\end{table}

\subsection{VCA Operators}

\sys defines a library of composition operators.  For space constraints, we will focus on its two core operators:
statistical composition, which explicitly computes the difference (or another measure) between matching marks in the two compared views,
and union composition, which spatially organizes the marks in the two views to aid visual comparison.  
In general, \sys focuses on the data transformations needed to define composition, and borrows the output visual mapping from its first argument.  

\subsubsection{Statistical Composition $\odot$}
\label{ss:algebra_stat}

$V^*=V_1\odot_{op} V_2$ joins rows from $Q_1$ and $Q_2$ and computes a new measure $Q_1.y\ op\ Q_2.y$ from the two views' measures (\Cref{f:vca}(a)):
\begin{align*}
  Q^* &= \pi_{A_{gb}, Q_1.y\ op\ Q_2.y\to y }\left(Q_1 \fullouterjoin_{A_{gb}} Q_2  \right) \\
  R^* &= R_1
\end{align*}
$Q^*$ first computes the outer join between $Q_1$ and $Q_2$ by matching records from each query 
whose grouping attributes  $A_{gb}$ have the same values.  
An outer join ensures that rows in either table have at least one output row even if there is no match.
Since the input queries, by definition, were grouped on $A_{gb}$, we are guaranteed exactly one output row for each group in $Q_1$ and $Q_2$.  Finally, $\pi$ copies the join attributes, computes $Q_1.y\ op\ Q_2.y$, and renames it as $y$.

$op$ is defined as ``$-$'' by default, however any binary arithmetic function is allowed.
As shorthand, $\oplus$ and $\ominus$ denote $\odot_+$ and $\odot_-$, respectively.
$\odot_{op}$ is symmetric iff $y_1\ op\ y_2 = y_2\ op\ y_1$.

%

%
\subsubsection{Union Composition $\cup$}

Union $V^* = V_1\cup_{qid,a}V_2$ composes the marks from both views into the same output view (\Cref{f:vca}(b)):
\begin{align*}
  Q^* & = \pi_{*, \hl{qid}}(Q_1) \cup \pi_{*, \hl{qid}}(Q_2) \\
  R^* &= \{qid\to\hl{$a$}\} \cup R_1 \hspace{2em} s.t. \ a \textrm{ is an unmapped vis attr}
\end{align*}
Each query $Q_i$ is augmented to track a unique identifier $qid$, so that rows from each query can be distinguished in $Q^*$.  $R^*$ additionally maps $qid$ to a visual attribute $a$ that is not already mapped in $R_1$.



\subsection{Safety}

Since \sys operators may take arbitrary combinations of marks as input, it is possible that the user may try to compose ``incomparable'' data together.  For instance, comparing stock price per hour with average stock price is sensible (safe), but comparing stock price with trading volume is not (unsafe).   

\sys defines safety based on the ability to map table $R_1$'s 
schema $A_{R_1}$ to table $R_2$'s schema $A_{R_2}$.  
Specifically, if their measures are the same, and if there is a 
unique mapping from each dimension in $A_{R_1}$ to a dimension 
in $A_{R_2}$ (or null).  
Dimensions can be mapped if they are the same.
In short, $A_{R_2}$ is a subset of $A_{R_1}$.

Although correct, this strict definition rejects many natural comparisons.
For instance, \sys cannot compare price per hour with price per day, because \texttt{hour}
and \texttt{day} are different dimension attributes.
Similarly, \sys cannot compare votes per district with votes per state.

\section{VCA for Hierarchical Data}

This section extends statistical and union composition to hierarchy-aware comparisons.
Given two views $V_1=R_1(Q_1(D))$ and $V_2=R_2(Q_2(D))$ and a hierarchy $H$,
we will define the semantics for statistical composition $\odot$ and union composition $\cup$.

\begin{figure}[htpb]
  \centering
  \includegraphics[width=.85\columnwidth]{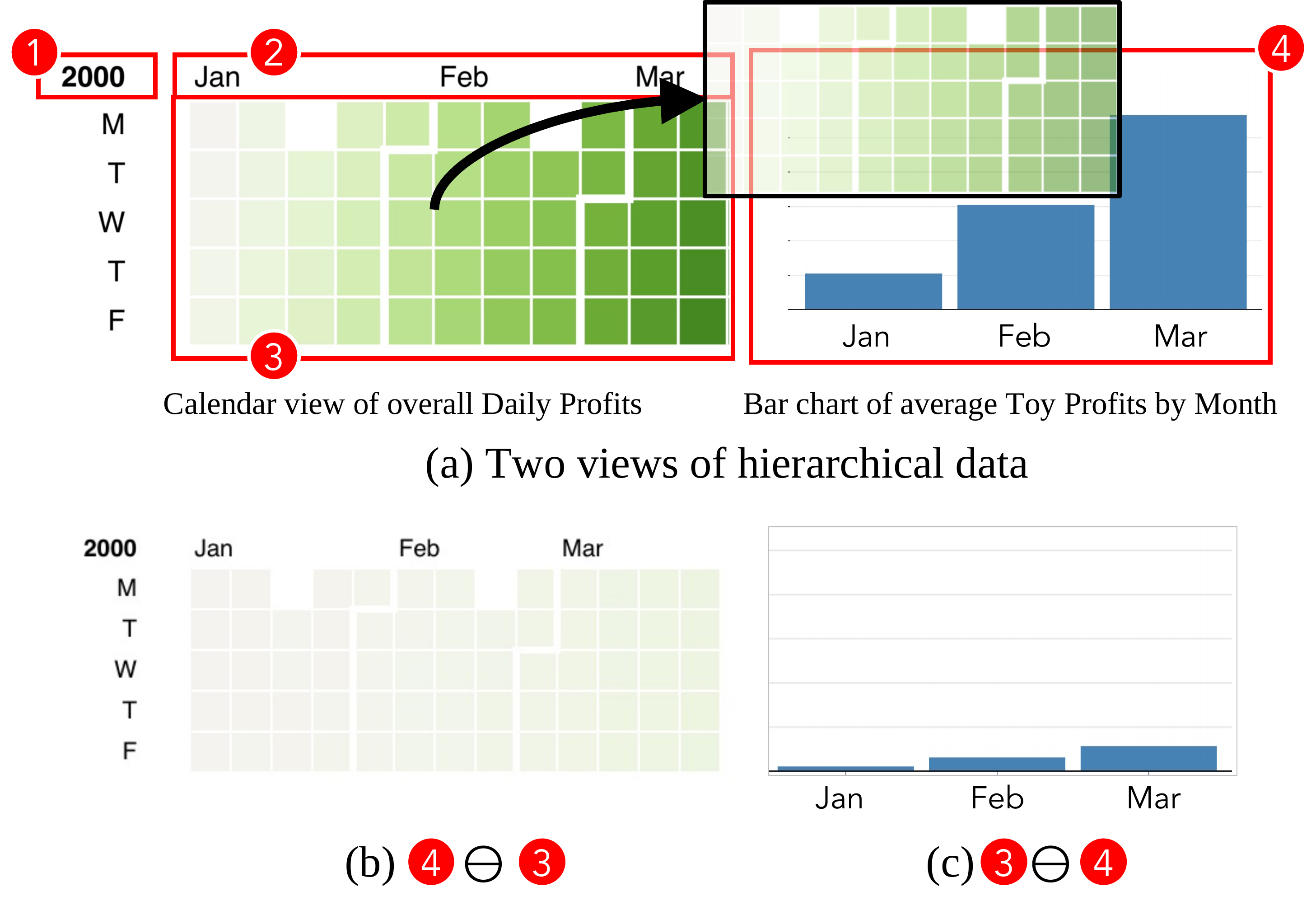}
  \caption{A hierarchical calendar visualization that renders daily profits, and a bar chart that renders average monthly profits.   All labeled subviews correspond to hierarchy levels that are comparable in \sysh. }
  \label{f:calendar}
\end{figure}

Our examples will be based on the visualizations in \Cref{f:calendar}, which builds on the hierarchy and data in \Cref{f:hierarchy}.  The calendar chart renders daily profits over all geographies and products, while the bar chart renders average monthly profits for toy products.    The goal is to define composition for any combination of comparisons between  visualization's labeled subviews (\circled{1} to \circled{4}).

\subsection{Safety}

To begin, we must extend \sys's safety rules to allow hierarchy-aware comparisons.
For space constraints, we focus on the case where both view's schemas ($A_1$, $A_2$) have the same number of dimensions,
as this is the core challenge.

$A_1$ and $A_2$ are safe to compose if there exists a bijective mapping:
$$\{ (a_i,a_j)  | a_i\in A_1, a_j\in A_2, a_i=a_j \lor \red{a_i\mapsto_H a_j \lor a_j\mapsto_H a_i} \}$$
Where $a_i\mapsto a_j$ given a directed path from $a_i$ to $a_j$ in $H$.
This relaxes \sys's rules so $a_i$ can map to $a_j$ if one is an ancestor of the other.

We also define $a_i{\sim}a_j$ to denote that the value of $a_i$ (e.g., "California") functionally determines $a_j$ (e.g., "USA") or vice versa.
If $a_i\mapsto a_j$, we can translate $a_i$ into $a_j$.
Consider $block\mapsto nation$ in \Cref{f:hierarchy}. We can translate block 1 to ``USA'' by 
joining $T_2$ and $T_3$ so that both attributes are in the same table, and then simply lookup 
the \texttt{Nation} attribute value.

\subsection{Statistical Composition $\odot$}

In \sys, $V^*=V_1\odot_{op} V_2$ matches rows in $Q_1$ and $Q_2$ using a join and computes a new measure $Q_1.y\ op\ Q_2.y$.  Extending $\odot$ to hierarchical data presents two challenges.  

First, it potentially requires reaggregating data in one of the views in order to compute the new measure.  
For instance, \circled{4}$\ominus$\circled{3} in \Cref{f:calendar}, which computes the difference between average monthly toy profits and overall daily profits.  The views are safe to compare, because \texttt{day}$\mapsto$\texttt{month}. However, we need to reaggregate the daily profits to the month granularity, and the specific aggregation function may depend on the application needs.    
For this reason, we extend statistical composition $\odot_{op,reagg}$ to take an additional $reagg()$ parameter for reaggregation.
$reagg$ defaults to $Q_2$'s aggregation function if not specified (e.g., $\odot_{op}$, $\ominus$).

Second, $\odot$ is asymmetric, and reaggregation is not always necessary nor desired.  
For instance, \circled{3}$\ominus$\circled{4} subtracts e.g., March's profits from each day's profits in March.   

For these reasons, we define $\odot_{op,reagg}$ under two cases based on whether or not reaggregation is required.
To simplify the discussion, we assume that $A_1$ is identical to $A_2$, except for a single pair of attributes $a_1\in A_1$ and $a_2\in A_2$, where $a_1\mapsto a_2$ or $a_2\mapsto a_1$.


\subsubsection{Case 1: $a_1\mapsto a_2$}

If the attribute $a_1$ is finer granularity than $a_2$, then we do not need to perform reaggregation.  
Thus we define the output view $V^*=V_1\odot_{op,reagg} V_2$ as follows:
\begin{align*}
  Q^* &= \pi_{A_{gb}, Q_1.y\ op\ Q_2.y\to y }\left(Q_1 \fullouterjoin_{joinp{\land Q_1.a_1{\sim}Q_2.a_2}} Q_2  \right) \\
  R^* &= R_1\\
  A'_{gb} &= A_{gb} - \{a_1, a_2\} \\
  joinp &= \land_{a\in A'_{gb}} Q_1.a = Q_2.a
\end{align*}
where $A_{gb}$ are the grouping dimension attributes in $Q_1$. 
{\it joinp} is identical to the join conditions in the definition in \Cref{ss:algebra_stat},
but only for the attributes that are identical in $Q_1$ and $Q_2$. 
The join also checks that the value of $a_1$ functionally determines the value of $a_2$.

\begin{example}
  Alice compares the difference between daily profits and monthly toy profits, so
  specifies \circled{3}$\ominus$\circled{4}, which results in \Cref{f:calendar}(b).
  The join condition only matches rows where translating $Q_1.day$ into its month is equal to corresponding $Q_2.month$.
  Each day's profit is reduced by that month's average profits. For instance,
  January 1st's profits is reduced by the average January profit.
\end{example}

\subsubsection{Case 2: $a_2\mapsto a_1$}

If the attribute $a_1$ is coarser than $a_2$, then we need to reaggregate the data in $V_2$.  Recall that queries have the form $Q_i = \gamma_{A^i_{gb}, f(a_y)\to y}(\sigma_{p}(D))$.
We define the output view $V^*$ as follows:
\begin{align*}
  Q^* &= \pi_{A^1_{gb}, Q_1.y\ op\ Q'_2.y\to y}(Q_1\fullouterjoin_{A^1_{gb}} Q'_2) \\
  Q'_2 &= \gamma_{A^{1}_{gb}, reagg(a_y)\to y}(\sigma_p(D)) \\
  R^* &= R_1
\end{align*}
The main idea is to reaggregate $Q_2$ at the same granularity as $Q_1$ (by using its grouping attributes $A^1_{gb}$), and then perform statistical composition as normal.  
We reaggregate the base table $D$ rather than the output of $Q_2$ to avoid potentially misleading statistics such as the average of averages.

\begin{example}
  Alice now compares the difference between monthly toy profits with the daily profits, so specifies \circled{4}$\ominus$\circled{3} (\Cref{f:calendar}(c)).
  To compute this, \sysh first aggregates e.g., March's profits using $avg$, and then subtracts the average profits from March's toy profits.  
\end{example}

%

\subsection{Union Composition $\cup$}
Although $\cup$ was a symmetric operator in \sys, it is asymmetric in \sysh.
This is because union composition requires two definitions based on the need for reaggregation.
Using the same notations and assumptions as above, we define $V^* = V_1\cup_{qid,a}V_2$ under two cases. 
$qid$ is a query identifier, $a$ is an available visual attribute.

\subsubsection{Case 1: $a_1\mapsto a_2$}

In the first case, $V_2$ is rendered as a coarser granularity than $V_1$,
so we duplicate each record in $Q_2$ for each matching record in $Q_1$.
For instance, $\circled{3}\cup\circled{4}$ will duplicate January's average profits for each day in January.
This is accomplished by defining $Q'_2$, which performs a left outer join, so that there is one output row for every unique value of $a_1$ in $Q_1$.  
We then keep the attributes in $A_1$ so that $Q'_2$ has the same schema as $Q_1$.
\begin{align*}
  Q'_2 &= \pi_{A_1}(\pi_{a_1}(Q_1) \leftouterjoin_{a_1{\sim}a_2} Q_2)\\
  Q^* & = \pi_{*, \hl{qid}}(Q_1) \cup \pi_{*, \hl{qid}}(Q_2) \\
  R^* &= \{qid\to\hl{$a$}\} \cup R_1 \hspace{2em} s.t. \ a \textrm{ is an unmapped vis attr}
\end{align*}

\subsubsection{Case 2: $a_2\mapsto a_1$}

In the second case, $V_2$ is at a finer granularity than $V_1$, so
we reaggregate its data to $V_1$'s granularity.  
This is similar to the reaggregation procedure for statistical composition, and also
requires a $reagg()$ function, which defaults to the aggregation function in $Q_1$.
\begin{align*}
  Q^* & = \pi_{*, \hl{qid}}(Q_1) \cup \pi_{*, \hl{qid}}(Q'_2) \\
  Q'_2 &= \gamma_{A^{2'}_{gb}, reagg(a_y)\to y}(\sigma_p(D)) \\
  A^{2'}_{gb} &= A^2_{gb} - \{a_2\} \cup \{a_1\}\\
  R^* &= \{qid\to\hl{$a$}\} \cup R_1 \hspace{2em} s.t. \ a \textrm{ is an unmapped vis attr}
\end{align*}

\section{Examples}\label{s:eval}

\sysh directly benefits from \sys's compositionality, its well-defined semantics, and that it can compare data independently of how they are visually encoded.
This section presents examples of how \sysh complements existing visual analysis systems.

\subsection{Tableau}

\begin{figure}[htpb]
  \centering
  \includegraphics[width=\columnwidth]{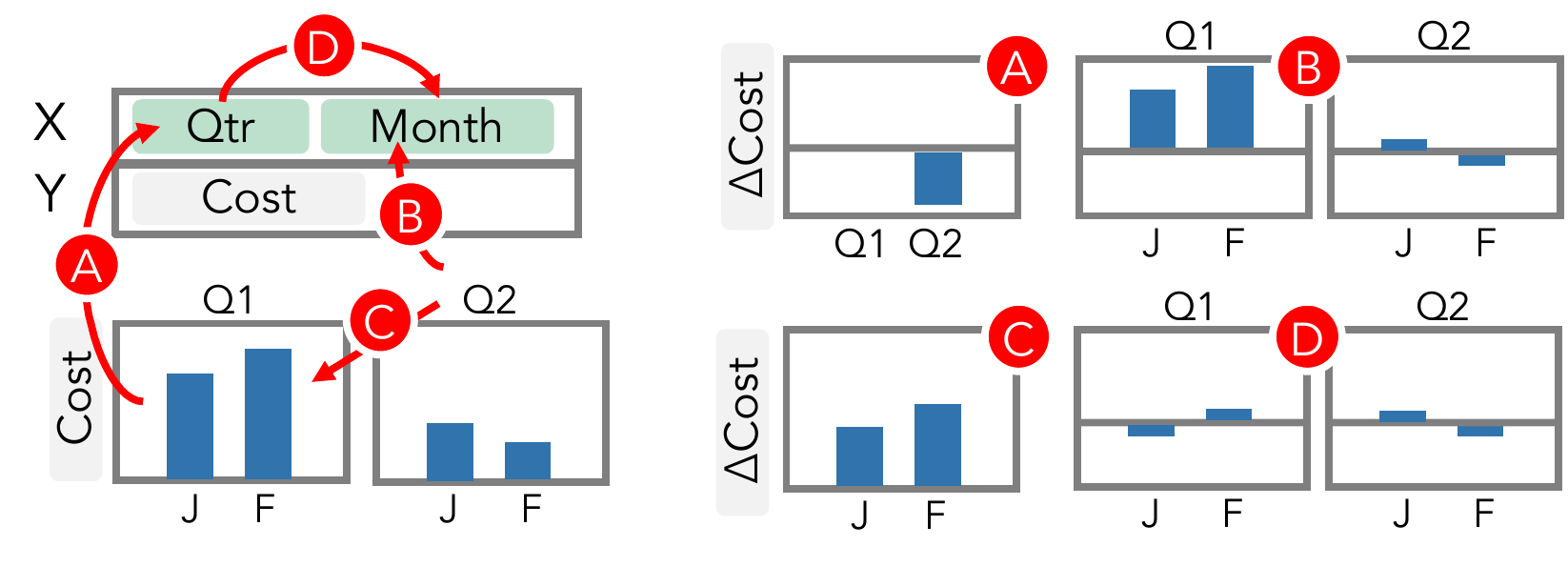}
  \caption{(left) Comparison interactions possible due to \sysh,  (right)
  the results of applying $\ominus$ to the interactions.}
  \label{f:tableau}
\end{figure}

\sysh can enable novel comparison interactions in Tableau-like exploration interfaces.
\Cref{f:tableau} shows small multiples of a bar chart for each quarter, and each chart renders costs per month.   
For all four example interactions, the user can drag component $V_2$ in the interface over $V_1$ to express $V_1\ominus V_2$. 

\circled{A} composes Q1's bar chart with all Quarters. It aggregates the bar chart data to the quarter level (average of Q1) and subtracts that value from each quarter's average cost.     The horizontal line is the zero line.
\circled{B} drags the Q2 header to the the month attribute in the x-axis shelf.  This subtracts  Q2's average cost from all small multiples charts.  The output is the same set of small multiples, but with each bar reduced by Q2's average cost.
\circled{C} drags the Q2 label onto the Q1 bar chart.  This is akin to \circled{B}, but only for a single target bar chart.
\circled{D} drags the \texttt{Qtr} attribute in the x-axis shelf onto the \texttt{Month} attribute. This computes each quarter's average cost, and subtracts it from the corresponding quarter's bar chart.   

\subsection{Hierarchical Visualizations in HIVE}

\begin{figure}[htpb]
  \centering
  \includegraphics[width=.8\columnwidth]{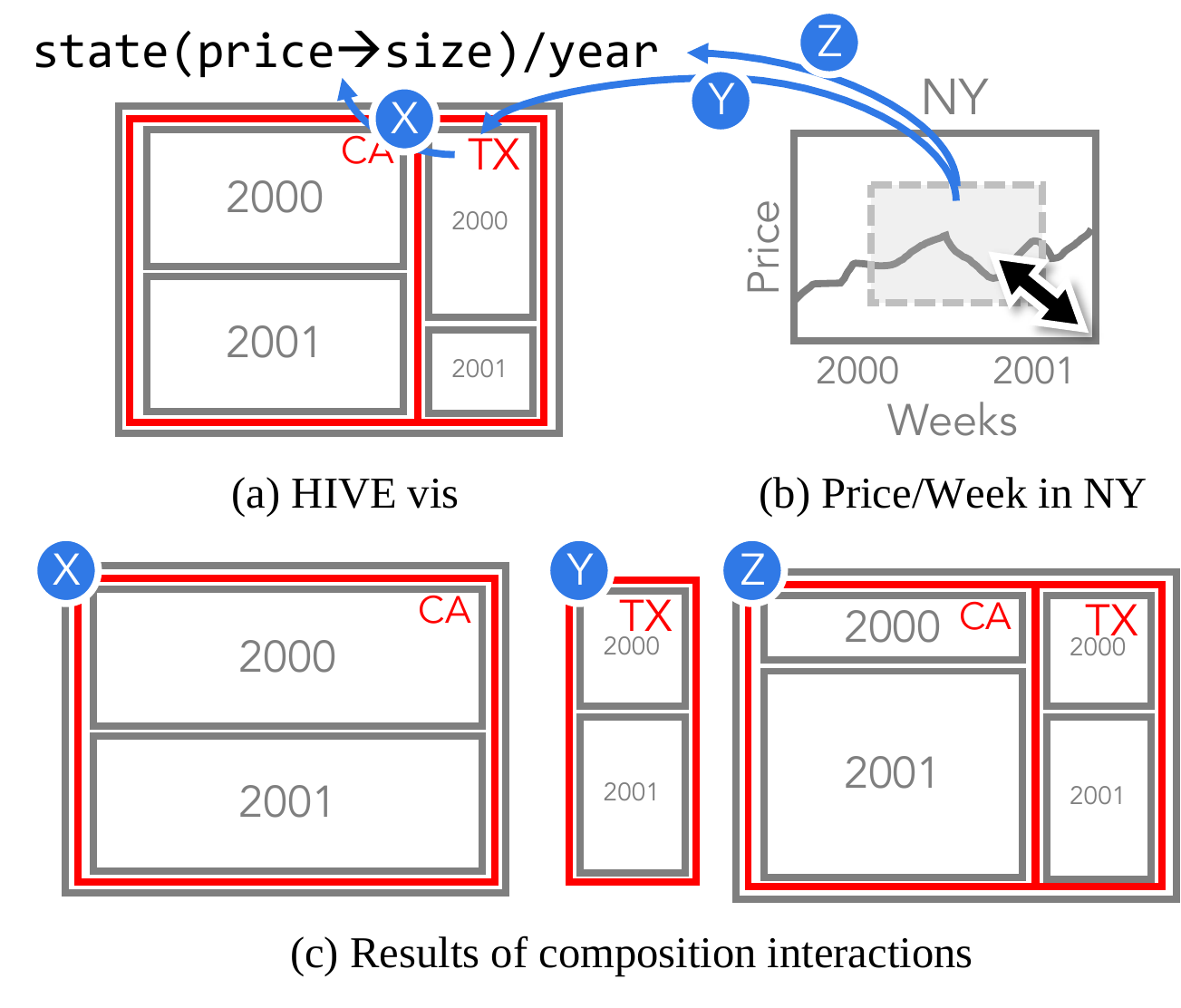}
  \caption{HIVE visualization hierarchically renders year within state, line chart renders price per week.}
  \label{f:hive}
\end{figure}

\Cref{f:hive} illustrates an example where the left 
view renders the HIVE statement \texttt{state(profit$\to$size)/year},
and the right view plots profit by quarter.
Logically, HIVE computes the spatial layouts one level at a time.
It first computes average price by state, and uses it to size the state-level \red{red} rectangles.
Then, for each state, it computes the average price by year and uses it to size the per-year rectangles.
Since these are all group-by aggregation queries, \sysh naturally applies.
The line chart renders price per week in NY, and \texttt{week}$\mapsto$\texttt{Year}.

We showcase three composition interactions and their outputs in \Cref{f:hive}(c).  
\bluecircled{X} drags Texas (\texttt{TX}) to the \texttt{state} component of the HIVE statement.   This will subtract Texas' 
average price from each state's price and re-render the entire spatial visualization.    Notice that TX disappears because
subtracted price is now 0, and is allocated no space.
\bluecircled{Y} drags a selection range in the line chart over the \texttt{TX} label.  This aggregates the data in the selection range to the year granularity and removes the aggregated values from the corresponding \texttt{TX} year.   Only the \texttt{TX} subset of the spatial visualization is returned.  
\bluecircled{Z} drags the selection range onto the \texttt{year} component of the HIVE statement.  This again aggregates the selected data, and subtracts it from each state's corresponding years.   The rectangles for 2001 are allocated more space in both states, however the proportions for the states do not change because the same amount is removed from both states.


\section{Conclusion}

View Composition Algebra (\sys) is a recent formalism for composing entire of parts of visualizations to aid adhoc comparison tasks.  Users can select values, marks, legend elements, and entire charts as targets, and use composition operators to summarize or compare the targets.  
This paper extend View Composition Algebra to support comparsions between views that render data at different hierarchical granularities.  This enables users to easily compare data within spatial visualizations, between different visualizations 
We presented the formal semantics that are compatible with the underlying queries supported by visual analytic systems such as HiDE and Polaris.   We further illustrated example interactions  in the context of small multiples visualizations and spatial hierarchical visualizations.

\bibliographystyle{abbrv-doi}
\bibliography{main}



\end{document}